 \documentclass[final,5p,times,twocolumn,authoryear]{elsarticle}

\usepackage{amssymb}
\usepackage{lipsum}
\usepackage{xcolor}
\usepackage{graphicx}
\usepackage{comment}
\usepackage{xurl}
\usepackage{csquotes}
\usepackage{array}
\usepackage{txfonts}

\journal{Astronomy $\&$ Computing}

\begin{document}

\begin{frontmatter}



\title{Learning from the present for the future: the Jülich LOFAR Long-term Archive}

\author{C.~Manzano, A.~Miskolczi, H.~Stiele\corref{cor}$^{\ast}$, V.~Vybornov, T.~Fieseler, S.~Pfalzner}
\ead{c.manzano@fz-juelich.de, a.miskolczi@fz-juelich.de, h.stiele@fz-juelich.de, v.vybornov@fz-juelich.de, t.fieseler@fz-juelich.de, s.pfalzner@fz-juelich.de}
\affiliation{organization={Jülich Supercomputing Centre, Forschungszentrum Jülich},
            addressline={Wilhelm-Johnen-Stra\ss e}, 
            city={Jülich},
            postcode={52428}, 
            state={NRW},
            country={Germany}}

 
\begin{abstract}
The Forschungszentrum J\"ulich has been hosting the German part of the LOFAR archive since 2013. It is Germany's most extensive radio astronomy archive, currently storing nearly 22 petabytes (PB) of data. Future radio telescopes are expected to require a dramatic increase in long-term data storage. Here, we take stock of the current data management of the J\"ulich LOFAR Data Archive, describe the ingestion, the storage system, the export to the long-term archive, and the request chain. We analysed the data availability over the last 10 years and searched for the underlying data access pattern and the energy consumption of the process. We determine hardware-related limiting factors, such as network bandwidth and cache pool availability and performance, and software aspects, e.g.\ workflow adjustment and parameter tuning, as the main data storage bottlenecks. By contrast, the challenge in providing the data from the archive for the users lies in retrieving the data from the tape archive and staging them. Building on this analysis, we suggest how to avoid/mitigate these problems in the future and define the requirements for future even more extensive long-term data archives.
\end{abstract}




\end{frontmatter}




\section{Introduction}

In everyday life, the total amount of data being stored grows exponentially \citep{data2017exponential}.\footnote{\url{https://www.cisco.com/c/en/us/solutions/collateral/executive-perspectives/annual-internet-report/white-paper-c11-741490.html}} 
The same trend transpires in science, including astronomy \citep{2006Natur.440..413S}, and future will produce orders of magnitude more data. Thus, the community must develop new concepts to process and store these vast amounts of data. Here, we concentrate on data archiving in this context. 

Currently, one of the most prominent example of data archiving in radio astronomy is the \emph{Low-Frequency Array} \citep[\emph{LOFAR};][]{2013A&A...556A...2V}.
The LOFAR \emph{Long-Term Archive} \citep[\emph{LTA};][]{2011ASPC..442...49R} is a distributed data archive. The data generated by the LOFAR radio telescope is pre-processed on a computer cluster in Groningen and then distributed later to one of the three sites which currently form the LOFAR Long Term Archive: the Samenwerkende Universitaire Reken Faciliteiten (SURF; Cooperating University Computing Facilities) in the Netherlands, the Jülich Supercomputing Centre (JSC) in Germany, and the Poznan Supercomputing and Networking Center (PSNC) in Poland.
Here we concentrate on the part of the LTA located at Forschungszentrum Jülich GmbH (FZJ), Germany. In this paper, we reflect on the lessons learnt from the current LOFAR LTA at FZJ to be prepared for the up-coming challenges arriving from up-coming telescopes like the Square Kilometre Array Observatory (SKAO), which will have much more extensive data production rates\footnote{\url{https://www.skao.int/en/explore/big-data}}.

LOFAR consists of an enormous array of omnidirectional radio antennas spread over many European countries, with the main hub being located in the Netherlands. Unlike in other area antennas, the signals from the separate antennas are not connected directly electrically to act as a single large antenna. Instead, the two types of LOFAR dipole antennas can be partly combined in analogue electronics, then digitised, and then combined again across the whole station. The data from all stations are transported over fibre to a central digital processor and combined in software to emulate a traditional radio telescope dish. The resolving power equals the greatest distance between the antenna stations.

Consequently, LOFAR is a radio interferometer with an innovative computer and network infrastructure that can handle huge data volumes. LOFAR started its science operations in December 2012. Several tens to hundreds of Terabytes of data are produced daily by over 50 LOFAR stations distributed across Europe. Even after processing at the LOFAR Central Processing in Groningen, several Terabytes of data need to be stored in the archive daily. Several computing centres share this task. The \emph{Jülich Supercomputing Centre} (JSC)\footnote{\url{https://www.fz-juelich.de/ias/jsc/EN/Home/home_node.html}} covers the German share of the LOFAR LTA storing raw and pre-processed data. Currently (October 2023), at JSC, the storage capacity is 1.5 PB on disk and 21.6 PB on tape, with a growth rate of 2 PB/year. The J\"ulich LOFAR Archive allows fast and efficient access to LOFAR data for the international radio astronomy community.

\begin{figure*}
  \begin{center}
    \includegraphics[width=1.0\textwidth]{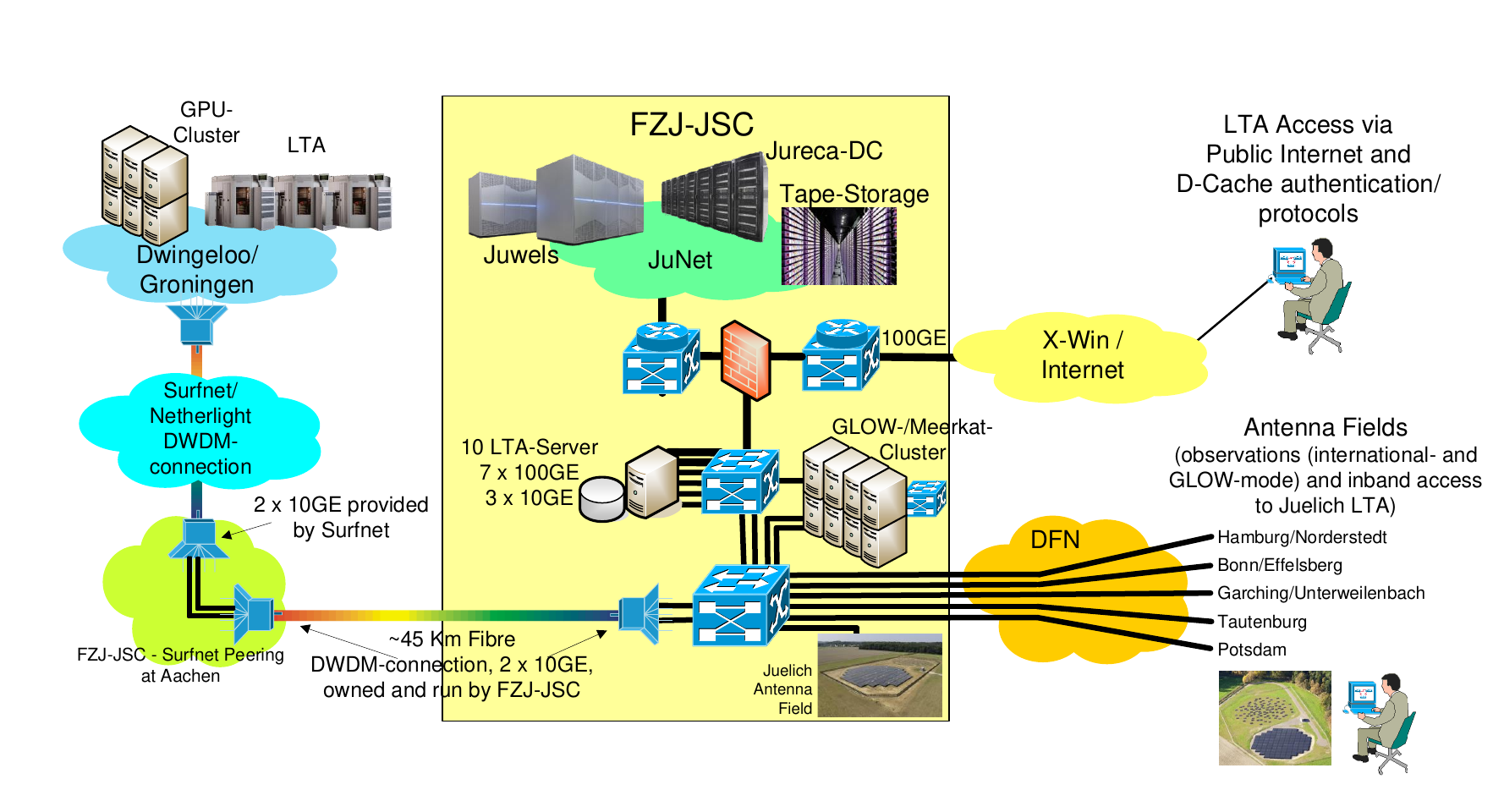}
 \caption{The JSC coordinates the data network activities of the GLOW stations. These are connected to the JSC via dedicated network links in a high-bandwidth star configuration. The real-time data from the antenna stations are transported from the JSC to local and international evaluation computer clusters using wave-division multiplexing technology. Pre-evaluated data then reach the Jülich LTA via the networks operated by the JSC, from where they are made available to the scientific community. (Copyright: Olaf Mextorf)}
  \label{fig:JSC_Network}
\end{center}
\end{figure*}

The six German LOFAR stations (GLOW stations)\footnote{\url{https://www.glowconsortium.de/lofar-about-new/lofar-in-germany-new/stations-in-germany/}} are connected round-the-clock to the LOFAR Central Processor in Groningen by fast glass fibres operating with 10-Gbit/s Ethernet. As gateway serves the JSC at FZJ, where the lines from the six \emph{GLOW} stations are combined onto two 10-Gbit/s lines (Figure~\ref{fig:JSC_Network}).

Because the network lines of all GLOW stations go via the JSC, all German stations can easily be connected via a shared virtual network (VLAN). Dedicated interface computers installed at the stations allow their control and data transport via the same VLAN. In principle, each station owner can operate several stations and store their data, limited only by the bandwidth of the network connections. Currently, the central control servers for managing the interface computers and for running the observations are located and maintained at the MPIfR Bonn\footnote{\url{https://www.mpifr-bonn.mpg.de/lofar}}. Data recording computers are hosted at the MPIfR Bonn and the JSC, allowing to receive data with total bandwidth from all GLOW stations in parallel. The GLOW mode, in which several of the German stations observe in parallel, is used mainly for pulsar observations\footnote{\url{https://www.glowconsortium.de/lofar-about-new/lofar-in-germany-new/single-station-science-new-2\#pulsarmonitoring}} .

In the following, we describe the current situation at the German LOFAR LTA. A concise overview of the storage system and data ingestion is given in   Sect.\ref{Sect:current}. The results of our analysis of the data life-cycle are presented in Sect.\ref{Sect:lifec}. Section \ref{Sect:Infra} contains the results of our infrastructure analysis, including the energy consumption. We present our reflections on current data handling in Sect.\ref{Sect:reflect}.
Based on these considerations we discuss next steps in long-term archiving, with respect to its ability to cope with the increasing demands of the next decade (Sect.\ref{Sect:next}).

\section{Current data management at the LTA}
\label{Sect:current}
\subsection{Criteria for data export to specific  archive}
The decision regarding which long-term archive will store the data from a specific observation is made by ASTRON and configured in the Telescope Manager Specification System  \citep[TMSS; see e.g.][]{10265351}. This decision is typically established per science project, considering the available storage capacity at the long-term archives. Additionally, considerations are made for the project's affiliations and data access/processing facilities of the team (e.g., users can express a preferred archive location when submitting a proposal for LOFAR observing time). Finally, service availability is important, especially if data needs to be moved from central processing and the preferred location is unavailable.

\subsection{Data ingestion into the Jülich LOFAR LTA}
Initially, the TMSS receives the beam-formed signals\footnote{Beamforming involves synthesising a directional sensitivity pattern to extract or emphasise signals coming from a particular region of the sky while potentially suppressing noise or interference from other directions.} from the different stations and triggers the pre-processing on the cluster in Groningen. At the end of the pre-processing the resulting data products, which mostly consist of visibility data, are ingested to remote LTA sites. 

Roughly 7 PB of data are ingested into the archive each year\footnote{\url{https://indico.skatelescope.org/event/551/attachments/5853/8647/9._LOFAR_-_RF_Pizzo.pdf}} and from those, approximately 2 PB are ingested at the Jülich LTA. The data is registered in a catalogue based on the Astro-WISE system \citep{2009MmSAI..80..509V, 2012Begeman_Astro-WISE} and made accessible through a web interface\footnote{\url{https://lta.lofar.eu/}}.

The ingest workflow includes performing checksum and tar operations. An entry is added as well in the LTA Catalogue for each data product. Once the data has been sent for long term archiving, the TMSS receives the storage location from the remote LTA and accordingly updates the corresponding catalogue entry.

\begin{figure*}
  \begin{center}
    \includegraphics[width=1.0\textwidth]{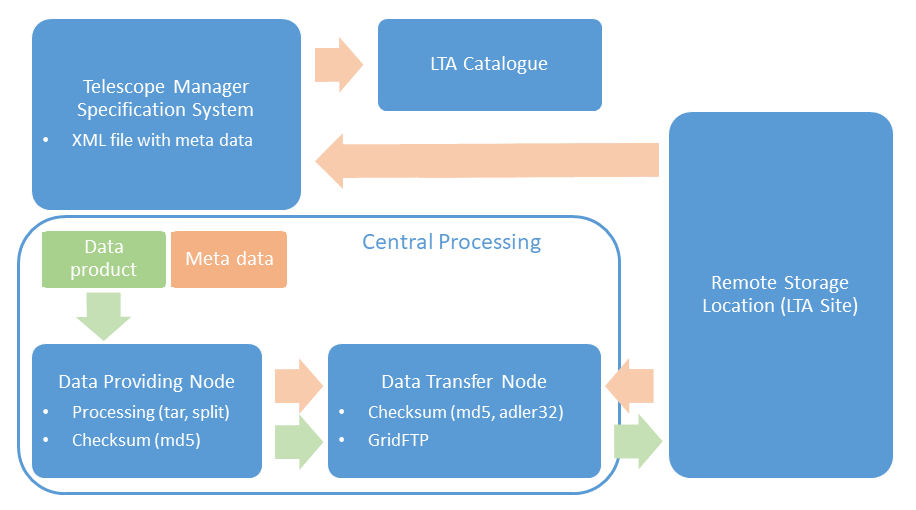}
 \caption{The figure shows the LOFAR ingest workflow. The data pre-processed on the central processing system is read by a Data Providing Node, subsequently processed, and then routed to a Data Transfer Node. From there, it is transmitted via GridFTP to one of the sites comprising the Long Term Archive. Throughout this process, data integrity is maintained through checksums and file size comparisons at various stages (for example, at the Data Transfer Node). Upon successful transfer to the Remote Storage Location, the Telescope Manager Specification System supplies the metadata, including observation properties and associated data products, to the LTA Catalogue.}
  \label{fig:LOFAR_Ingest_Workflow}
\end{center}
\end{figure*}

Regarding the LOFAR ingest workflow, it involves multiple steps (Figure~\ref{fig:LOFAR_Ingest_Workflow}).

The TMSS gathers metadata from the data products and compiles it into an XML file. This XML file is then utilised for exchanging information with the LTA systems. On the central processing system, there are data-providing nodes and data transfer nodes which take care of different tasks.

On the data-providing nodes, a script operates to read data from the disk, package multi-file data products into a single tar file, and perform an MD5 \citep{rfc1321} checksum calculation on the data. MD5, developed by Ronald Rivest in 1991, is a widely adopted hash function that produces a 128-bit hash value, serving as a checksum to ensure data integrity against unintended alterations. The data, along with the calculated checksum, are then streamed to a data transfer node.

On the data transfer nodes, the data stream received undergoes processing through a custom-built system that calculates various checksums, including Adler-32 \citep{rfc1950} and MD5. Adler-32, designed by Mark Adler, is a straightforward checksum algorithm derived from Fletcher's checksum, which is an error detection method utilising bit addition and weighted positions in the message \citep{DBLP:journals/tcom/Fletcher82}. It can identify errors, especially in compressed data streams with random transmission errors. The computed MD5 checksum value from this process is then cross-referenced with the value computed on the data-providing node for verification.

The data stream received by the data transfer node is piped into a GridFTP client. GridFTP allows sending data files over multiple TCP (Transmission Control Protocol) streams using multiple sockets across the same network path, maximising its throughput.  It handles the transfer to a dCache\footnote{\url{https://dcache.org}} instance on the remote LTA site.
At FZ Jülich, the data is sent to the archive via a dedicated private network and ingested in the Jülich LOFAR LTA. The remote site will automatically trigger an Adler-32 checksum calculation and include the resulting checksum in the file metadata within the local dCache file catalogue.

If the transfer succeeds, the data transfer node requests the checksum and file size from the remote site and verifies those numbers. In case of verification errors, the transfer will be automatically retried several times. In case the failure persists, the ingest will be marked as failed and will require operator intervention.

After successfully completing the data transfer, the metadata is submitted to the LTA catalogue. It undergoes validation, and if the validation is successful, the metadata is integrated into the catalogue database. With this registration process completed, the data becomes accessible for queries.

\subsection{The underlying system of the data storage}
The Jülich LOFAR LTA consists of a distributed heterogeneous storage system based on the dCache software. dCache allows storing and retrieving huge amounts of data distributed among many different server nodes under a single virtual file system tree with various standard access methods.

This hierarchical storage solution is made up of spinning disks for “online” or “hot” data and tapes for “offline” or “cold” data. The data ingested to the archive is first copied to the disk pools (cache) allowing fast ingestion (thus maximising bandwidth use), and asynchronously migrated to tape in a background process. A read request will trigger a process that stages the data again to cache (if not already there). The data is available for reading as long as it is in cache. The software interface ENDIT\footnote{\url{https://github.com/neicnordic/endit}} takes care of the communication between disk and tape. In the Jülich configuration, it allows the communication with an IBM Storage Protect\footnote{\url{https://www.ibm.com/products/storage-protect}} server which enables and handles the access to the tape libraries. This approach offers both long-term archiving on tape and efficient access to the international team of researchers for processing the data.

\subsection{Data Access}
Currently there are data requests from about 100 distinct users each year. LOFAR users who wish to access data in the LTA can first visit the LOFAR web service\footnote{\url{https://lta.lofar.eu/}}. Here, they can select specific data of interest, such as particular observations. Once in the archive view, users can request that the data is to be staged. This request is managed by a custom-built staging service, which handles staging requests for authorised users. The process initiates a request at the respective sites (in cases where the data is distributed across multiple sites of the LTA) containing the desired data. In Jülich, this request is forwarded to the dCache management server, triggering the ``bring online" process. All of this occurs in the background, unbeknownst to the user, who will receive an email notification only after the data has been successfully brought online. The email includes the download links for the requested files.

Data retrieval is a three step process. First the user compiles a list of files to be staged. This can either be done through the aforementioned ASTRON's LOFAR web service or a custom script where the SRM\footnote{https://sdm.lbl.gov/srm-wg/} protocol can be used to get a list of available files.
The second step is sending the staging request for the selected files. This is again be done via the LTA portal or a custom scrip. 
The third and final step is downloading the data. This can be done through two possibilities. 
The first one employs the gfal2\footnote{https://github.com/cern-fts/gfal2} software package which can download files using multiple protocols. This approach is mostly used by users that process data on HPC systems hosted at the Jülich Supercomputing Centre. The second approach is more general where the data is downloaded using common software that can download files using the http protocol.
Authentication for all steps is done either via x509 certificates or by using a VOMS\footnote{http://www.eu-emi.eu/products/-/asset\_publisher/z2MT/content/voms} certificate that is verified by the ASTRON VOMS server.
For downloads we also tested and implemented a third option which employs the WebDav protocol together with so called macaroons\footnote{https://ics.uci.edu/\textasciitilde{}ejw/papers/dav-ecscw.pdf} as authentication tokens. A macaroon is a string of alphanumerical characters in which multiple permissions are encoded. These permissions can contain, for example, path limitations or a certain time range in which these are valid. This authentication method is supported by WebDav and can also be used with common downloading software.
Once these steps are taken the time between between the request only depends on the network speed and the workload of the pools. If pools are available and the requested file is not on a tape that is currently used, the file can be ready within a few minutes. If the pools are busy with many active staging requests this can take hours. In extreme cases it can even take up to a day or two until the requested files are available to download. Contributing factors to this are pool availability, current network usage, tape/drive availability, time to insert the tape into the drive and seeking to the desired file on tape.

A detailed user documentation how to request data sets is available online\footnote{\url{https://www.astron.nl/lofarwiki/doku.php?id=public:lta_howto}}.

\subsection{Data availability}
\label{subsec:data_avail}
Once the data has been fed into an LTA site, there is currently no further active data management by the telescope management system and there is no need to move data between LTA sites.

The primary copy of the data is stored permanently on tape. When data is ingested or retrieved, it is copied in the cache (disk pool), where it gets a lifetime (it is "pinned"), which is 7 days by default.

From the available pools, dCache determines the pool used for storing or reading a file by calculating a cost value for each pool. The pool with the lowest cost is used. The total cost is a linear combination of the performance and space cost. The performance costs describes how "busy" a pool is, and the space cost describes how much it “hurts” to free space on the pool for the file. This last mentioned cost depends on the free space on the pool or on the age of the least recently used (LRU) file, which would have to be deleted. There are several parameters that affect this cost calculation and that can be tuned, for example, the gap parameter. Its default value is 4 GiB and it is the size of free space below which it will be assumed that the pool is full and consequently the least recently used file has to be removed. The recommendation is to set it to the size of the smallest files which frequently might be written to the pool.\footnote{\url{https://www.dcache.org/manuals/Book-9.2/config-PoolManager.shtml}}

The files deleted from disk must have a copy on tape and not be pinned.

\section{Data Life-cycle Overview}
\label{Sect:lifec}
\subsection{Statistic on data access}
\begin{figure*}
    \centering
    \includegraphics[width=0.9\textwidth]{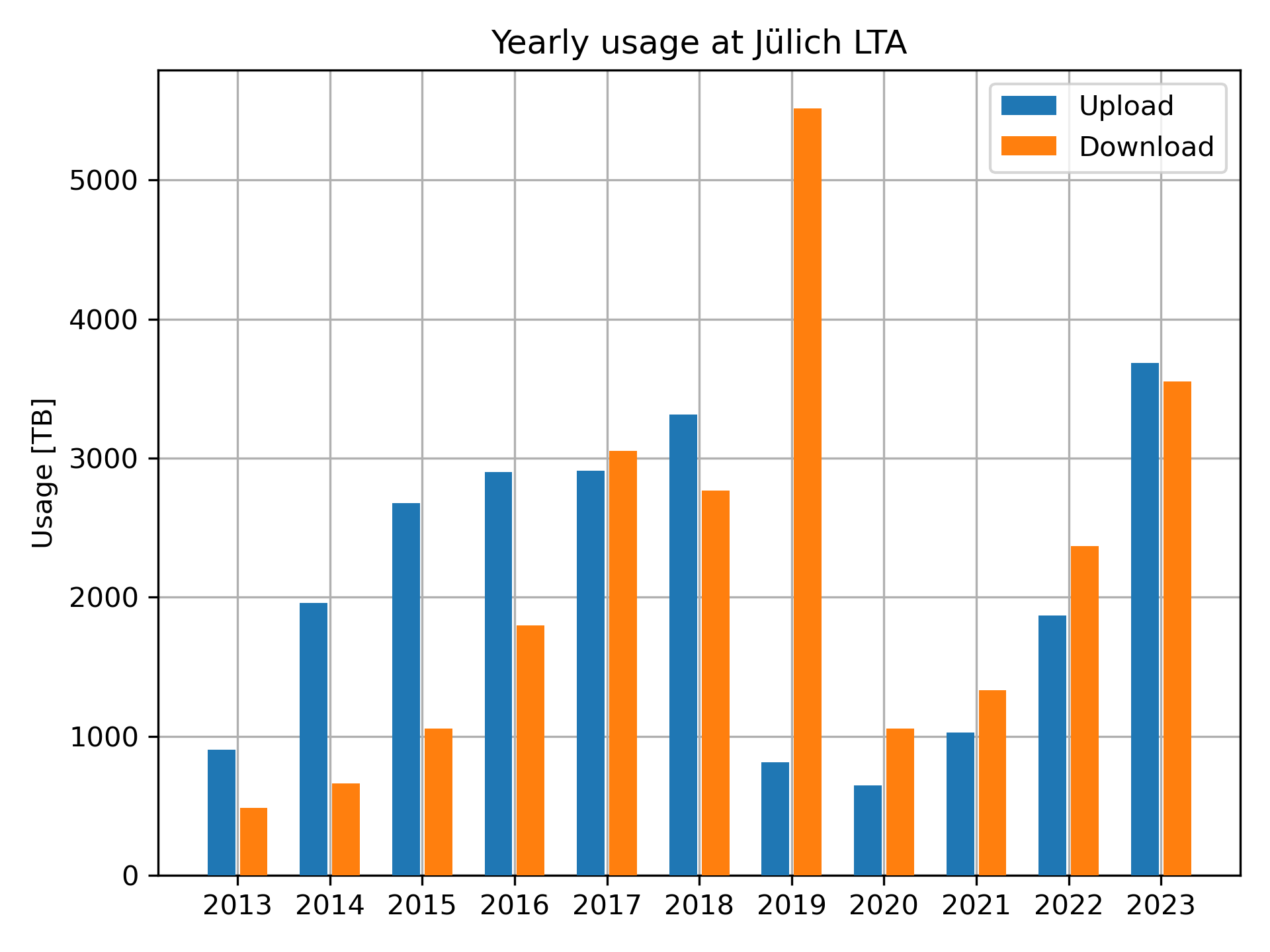}
    \caption{Up-, and Download statistics of the Jülich LTA from the beginning of 2013 to the end of 2023}
    \label{fig:Total_LTA_usage}
\end{figure*}

As of October 2023, the archive for LOFAR data at Jülich stores about 21.6 PB of raw and pre-processed data from the LOFAR telescope. Data acquisition started in late 2013. An agreement between JSC and ASTRON was made, which allowed ASTRON to upload 2 PB of data per year. This is not a hard limit. A deviation in both directions is accepted, which happens every year. At the same time, the amount of retrieved data is about the same, though in the first few years, it was less than the amount of uploaded data. This is due to the number of scientist working with the data has increased significantly over the last years. As figure \ref{fig:Total_LTA_usage} shows, in the first years, the amount of uploaded data adhered to the 2 PB/year, sometimes being below and above it. In 2019 ASTRON introduced DYSCO \citep{2016A&A...595A..99O} compression of LOFAR data which drastically reduced the size of the data sets \citep{2019A&A...622A...1S}. This is reflected in the drop in uploaded data in 2019. The jump in downloads in 2019 is due to the LoTSS project \citep{2019A&A...622A...1S}, for which bulk processing began in that year.

\subsection{Details of stored files}
\begin{figure*}
    \centering
    \includegraphics[width=0.9\textwidth]{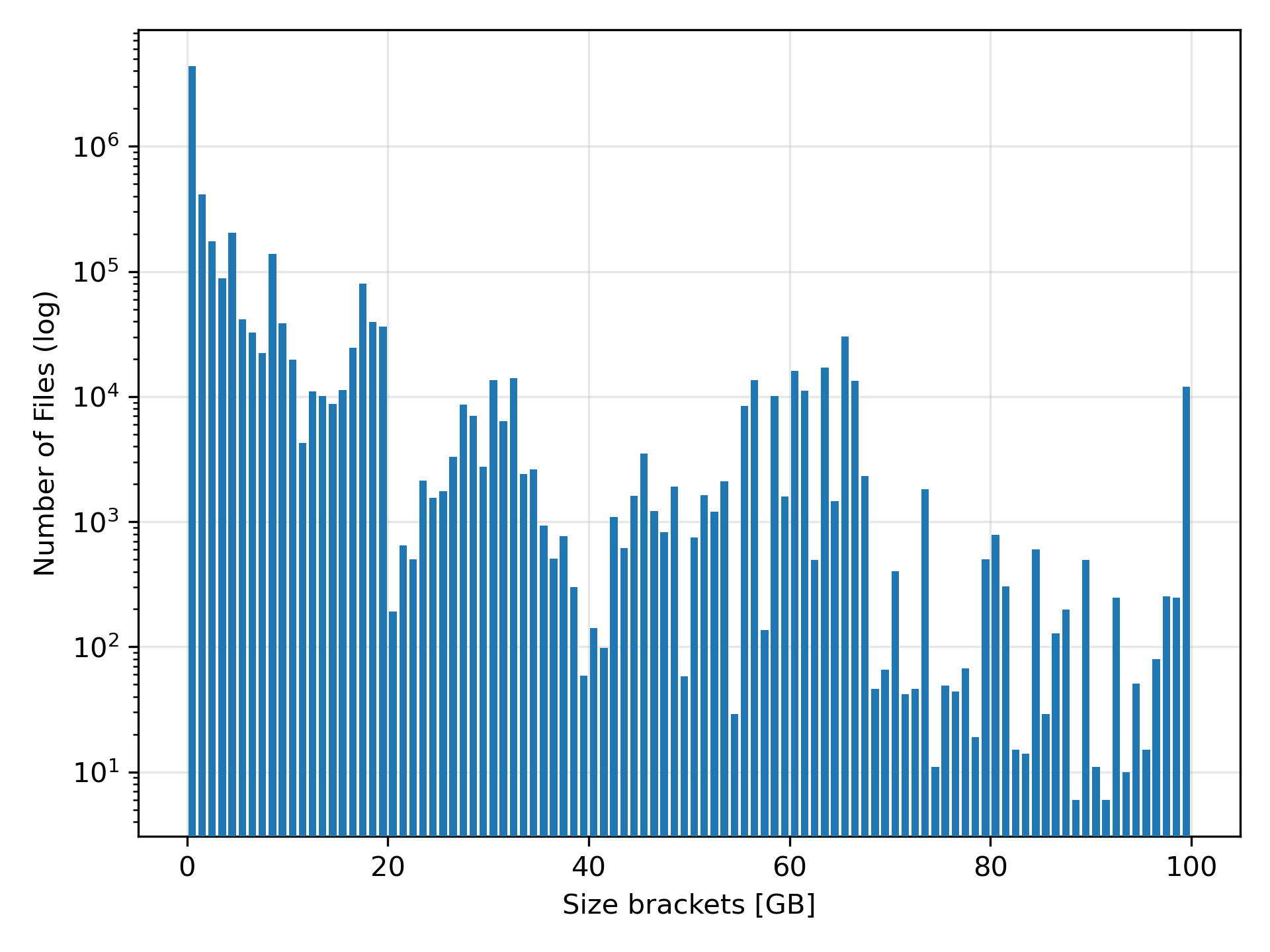}
    \caption{File size distribution of stored LOFAR files as of December 2023. The number of files is shown on a logarithmic scale. Files equal to or larger than 100 GB are added together for easier readability.}
    \label{fig:Total_file_distribution_all}
\end{figure*}

Currently, the Jülich LTA stores about 5 million files. The size distribution of the files is shown in figure \ref{fig:Total_file_distribution_all}. On the x-axis, the file sizes are shown in brackets of 1 GB. The smallest files are a few bytes small, they are mostly old test files.
The largest file stored has the size of 469 GB. Since there are only 514 files equal to or larger than 100 GB, they are summed up for easier readability of the figure.
The 1 GB bracket has by far the most number of files with $\approx 3.9$ million files, accounting for $\approx 76\%$ of all files. Most of these files are calibrator observations.

\subsection{Data access pattern for downloaded data}
\begin{figure*}
    \centering
    \includegraphics[width=0.9\textwidth]{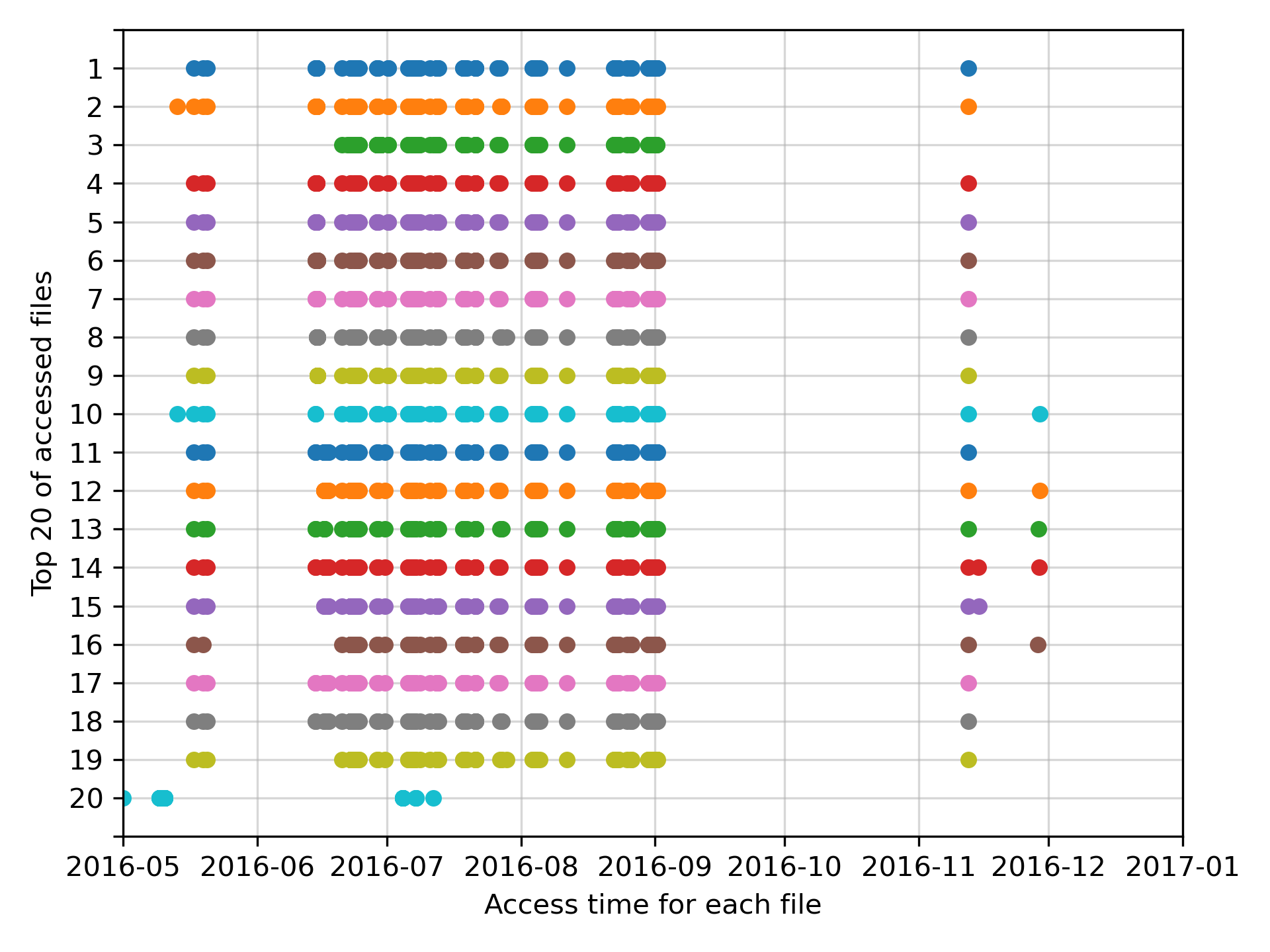}
        \caption{Top 20 files which were downloaded the most times. The cumulative number of downloads for each file is plotted against the times of access. A clear pattern emerges: Many downloads in the first three months, followed by two month gap. A single file is represented by one colour (but please note that colours have to be reused to indicate files with different cumulative numbers of accesses). }
    \label{fig:Top20Files}
\end{figure*}

Currently, there is no download counter that influences the time that the data is available. To consider the usefulness of such a mechanism, we analysed the data access of the LOFAR data. Therefore, we selected the top 20 files with the most accesses and plotted them on a time-vs.-access plot as shown in \mbox{figure \ref{fig:Top20Files}.}

Only files were selected that had a complete transfer not to be influenced by accesses that repeatedly tried to continue or retry a broken transfer. Regularly downloaded test files that are tested every day are also omitted.

An access pattern emerges for those files which are downloaded the most. All of the top 20 files are from mid-2016, a time where multiple processing programs and pipelines were developed and tested. The pattern indicates that shortly after data is ingested into the archive, users start downloading the data for early processing for the next three months. After that time frame only random access occur which can not be accommodated for each file.

The way dCache works, a file can be pinned for a certain amount of time. In that time dCache makes sure that the file is immediately available for downloads without retrieving it from the tape library first. By default, this time is only a week long (see also Subsection \ref{subsec:data_avail}). That only means that the file \textit{can} be overwritten if the space is needed for new files, but it does not mean that it will be deleted after the time has passed. Due to this mechanism, files can be accessed many weeks afterwards, but depending on the amount of uploaded data, they can be deleted on the cached disk space.
Currently, the Jülich LTA has about 1 PB available as cached disk space. In normal operations, files that are uploaded to the Jülich LTA are first stored on the cached disk and then immediately copied to magnetic tapes. This can take a while since our libraries are shared with many projects, and sometimes the transfers must wait until drives are available. 

Assuming an ingest amount of 2 PB per year, in those 3 months, roughly 500 TB would be uploaded. If the pinning time of each newly uploaded file were set to three months to make sure that users can immediately download them without any wait time, there would be a low risk of running out of space for new uploads.
Therefore, keeping files available longer would be viable on the Jülich LTA in its current state.

\section{Infrastructure Analysis}
\label{Sect:Infra}
\subsection{Storage of final data products}
\label{subsec:Storage_of_final_products}

Depending on the type of computation or data it might make sense to use HPC resources for processing the LOFAR LTA data (see also Subsection \ref{subsec:Reflection_on_data_volumes_and_computing_time}). An example of this is the processing of the LoTSS data \citep{Drabent:874406}. The survey consists of 3168 pointings and requires about 30 PB of storage space. About 50\% of all LoTSS observations conducted to date are stored in the Jülich LTA, i.\,e.\ 800
observations occupying about 12 PB. Such an amount of data has to be processed in an automated, well reproducible and organised way within a reasonable time. Efficient processing therefore requires computing facilities with a fast connection to the data storage in order to avoid unreasonably long transfer times of the data to local computing resources via the Internet.

In Jülich HPC resources are available for post processing, and in order to process data the scientist needs to be member of a compute project. As there is no direct access to the dCache system on the compute nodes, once access to the resources is granted, to compute on the Jülich LTA data, the user has to copy them over to the disk storage of the compute project. This can be done through the Jülich Data Access (JUDAC)\footnote{\url{https://www.fz-juelich.de/en/ias/jsc/systems/storage-systems/judac}} servers which serve as a gateway to the facility-wide file systems and have as well a fast connection to the dCache systems in Jülich.

However, after the computation, final data products such as radio maps or spectra are presently not stored within the Jülich LTA, and the user has to take care of these outcome/results and bears the sole responsibility for preserving copies of their results. This solution is suboptimal, as valuable data could potentially be lost under this setup.

\subsection{Energy consumption}

The Jülich LTA consists of different components: servers, disk servers, tape libraries, network elements like switches, etc. At the moment, it is impossible to obtain detailed live values for the different components other than aggregated ones. Also, some of the components are shared resources like the tape libraries or some of the network components. However, it is possible to make an estimation of the energy consumption of the Jülich LTA setup.

A regular server used in the setup consumes, on average, 144 Watts\footnote{In our case, this value can be obtained from the Integrated Dell Remote Access Controller interface.}. Regarding the disk servers, taking into account the disk consumption values provided by the vendor\footnote{See for example the data sheet from a regular disk used in our setup: \url{https://www.seagate.com/www-content/datasheets/pdfs/exos-x-14-channel-DS1974-4-1812US-en_CA.pdf}} we can estimate power consumption of 148,8 Watts per disk server\footnote{A disk server consists of a controller and 24 spinning disks. We disregard in this calculation the power consumption of the controller}. Considering these numbers and the number of servers and disk servers, we can evaluate a power consumption of roughly 23000 kWh/year. Because these are the components used for the system's cache, we can also calculate from this a power consumption of 13310 kWh/year/PB\footnote{We had at the moment of the calculation 1728 TB of cache capacity (disk).}.

Regarding the tape libraries, the LOFAR data stored in the Jülich LTA is distributed in 3 different tape libraries at the JSC facilities\footnote{The tape libraries are an IBM TS4500 and two Oracle SL8500 in two different buildings. These are shared resources available to all supercomputer users and users from the Forschungszentrum's campus.}. From these, we have different power consumption values provided by the vendors\footnote{The average power consumption of the two Oracle libraries is 3095 and 4607 Watts. The power consumption of the IBM library at 50\% load is 1262 Watts.}. Taking into account the percentage of LOFAR data regarding the total usage in each library, the power consumption is 151 Watts in the IBM library and 992 Watts combined in both Oracle libraries\footnote{At the moment of this calculation, 11\% of the total data stored in the IBM library was LOFAR data and 32\% of the total data stored in both Oracle libraries was LOFAR data.}. For the total amount of LOFAR data stored, this amounts to 536 kWh/year/PB\footnote{At the moment of the calculation, there were 18,7 PB of LOFAR data stored on tape.}.

The power usage of the shared network components is estimated to be negligible in comparison. 

\subsection{Monitoring}

In such a complex system, monitoring plays a critical role in ensuring proper functioning. When it comes to hardware, there are already mechanisms in place to trigger alarms when values deviate from expected ranges (e.g., CPU and memory usage) or when disks in the cache pools fail and need replacement. Various aspects of operations are also monitored, such as tracking the number and total size of ingested or downloaded files within a specific time frame, as well as cache pool utilisation (files already migrated to tape and pending ones). Additionally, other metrics like ping time and the number of threads used by the different services are monitored. Custom scripts are deployed to raise alarms when errors or unexpected behaviours are logged. Regular functionality checks include verifying the correct ingestion of test files into the archive and assessing network capacity through iperf tests.

\section{Reflection on current data handling}
\label{Sect:reflect}

\subsection{Bottlenecks for storing data}

The data storage is limited by factors from various sources, including hardware, software, and the human factor, all of which can impact the efficiency of data ingestion.
Hardware-related limiting factors include:

\begin{itemize}
\item Network bandwidth: The speed at which data can be ingested is constrained by the limitations of the network bandwidth. Higher network bandwidth allows for faster data transfer and ingestion.

\item Cache pools: The size and write speed of the cache pools receiving the data play a crucial role. The cache pools need to efficiently migrate ``cold" data to the back-end storage to make space available for incoming data. The speed of this migration process is vital to maintain smooth data ingestion.

\item Availability of hardware: Hardware maintenance slots are essential for ensuring the smooth functioning of the system. However, during these maintenance slots, service availability might be partially affected. To mitigate this potential problem, redundancy measures are usually in place, such as running multiple cache pools in parallel. While maintenance is typically planned and announced in advance, unforeseen hardware failures can still impact cache space availability during peak ingests. In addition, implementing strategies such as data replication across geographically dispersed nodes or using fault-tolerant architectures can minimise downtime and ensure continuous access to cache space even during maintenance or hardware failures.
\end{itemize}

\noindent
Regarding software aspects, limiting factors include:

\begin{itemize}
\item Software bugs: Although the software is thoroughly tested in advance, there can be instances where new software versions introduce unexpected side effects. These bugs may significantly affect the performance of data ingestion. Each use case may have its unique requirements, and the software updates may not have accounted for all the singularities, leading to undesired consequences. Maintaining a robust integration environment and implementing continuous integration/continuous deployment (CI/CD) techniques are essential to mitigating the risk of introducing bugs. By continuously integrating code changes into a shared repository and automating the deployment process, teams can identify and address potential issues early in the development cycle, ensuring smoother transitions between software releases and minimising disruptions to data ingestion processes.

\item Adjustment of workflow and parameter tuning: Optimising the system's performance for a specific use case often involves adjusting the workflow and tuning various parameters. However, such adjustments can have unintended consequences in other areas and for other use cases. For example, optimising cache pools for maximising staging throughput may adversely affect data ingestion throughput and vice versa. Balancing these trade-offs and finding the right configuration can be challenging.
\end{itemize}

\noindent
The human factor also plays a significant role in data ingestion and introduces its own set of limiting factors:

\begin{itemize}
\item Human factor: Misconfigurations, miscalculations, and errors in handling the system can impact data ingestion efficiency. Human errors, whether in configuring the system or making calculations, can lead to sub-optimal performance or even system failures. It is essential to have well-documented and well-communicated procedures in place to minimise the occurrence of such errors and provide robust error-handling mechanisms.
\end{itemize}

In summary, the limiting factors for efficient data ingestion encompass hardware-related aspects like network bandwidth and cache pool performance, as well as software-related factors such as software bugs and the delicate balance of workflow adjustment and parameter tuning. Addressing these factors requires careful consideration and testing to optimise the system's performance across various use cases. Additionally, misconfigurations, miscalculations, and error handling are crucial aspects that need as well to be taken into account.

\subsection{Bottlenecks for providing data}
Besides the factors mentioned in the last section which can have a bad impact on an efficient data workflow, also the data processing schedule can cause avoidable delay. As the current policy does only guaranty that data is stored in the disk cache for one week, a request which comes after this time has expired may result in triggering a retrieval from tape, which is far more expensive in terms of access time.

Staging data from tape requires the use of tape drives; these devices mount and read the data on magnetic volumes. Tape drives are a shared resource in Forschungszentrum Jülich, and they can be used by supercomputer users, backup and archive services, etc. In total four tape libraries are available at JSC\footnote{\url{https://apps.fz-juelich.de/jsc/hps/just/configuration.html#system-name-tape-libraries-automated-cartridge-systems}} each providing 20 to 38 tape drives. The availability and allocation of tape drives become critical in providing timely access to data. The limited number of tape drives, combined with their shared usage across different services, can potentially introduce delays in data retrieval and staging processes.

\begin{figure}[h]
    \centering
    \includegraphics[width=0.5\textwidth]{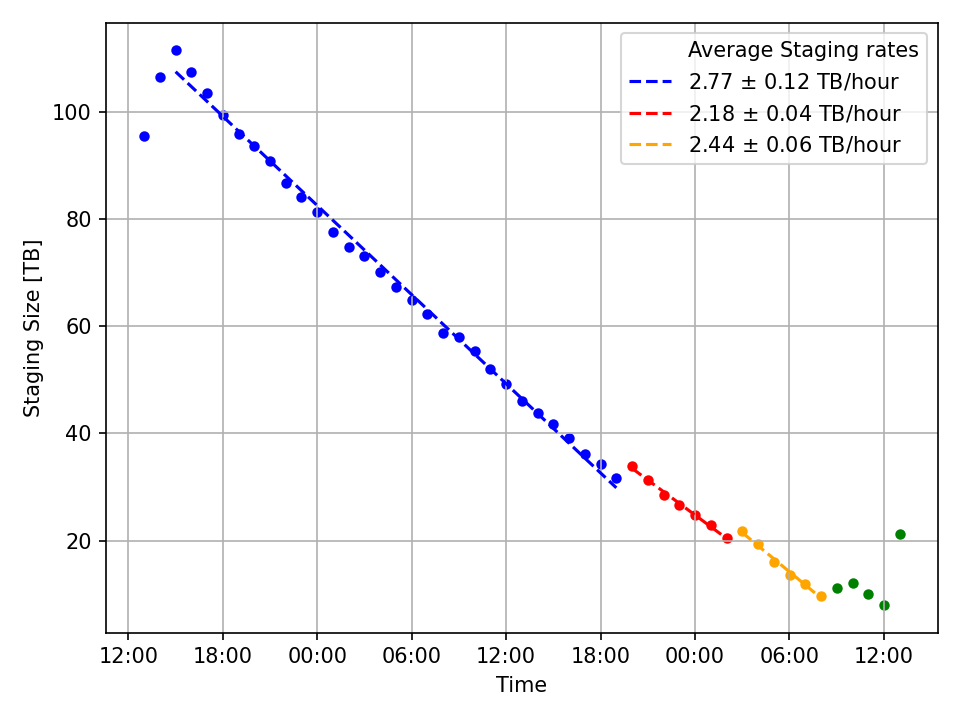}
    \caption{Staging rate in TB/hour over a 48 hour period of time for three different staging batches.}
    \label{fig:stagingrate}
\end{figure}

A second bottleneck in staging data arises when less tapes than pools are used. One tape can always only be used by one pool. If staging requests happen to have the files on one tape, logically, only one pool can retrieve them. Therefor, if the number of requested tapes is lower than the number of pools, not all pools will be actively getting data from the tape storage. In those cases the staging throughput is limited by the number of pools and the network throughput to each pool.
Figure \ref{fig:stagingrate} shows a 48 hour period from 22.04.2024 13:00 to 24.04.2024 13:00 where all pools were actively staging. The first two and last 5 hours were not used for the statistics due to the high number of additional requests, which would heavily skew the staging rate calculation. The average staging rate is 2.77 $\pm$ 0.12 TB/hour for the first 26 hours. During this time all pools had a tape to work. After this batch, two pools stopped working. Although there was enough data to keep all pools busy, the way the staging software works, the requests can not easily be redistributed. So for the next 7 hours two pools were not used, and this is reflected in a reduced staging rate of 2.18 $\pm$ 0.04 TB/hour. After this a new batch of requests arrived which again were distributed to unused pools, again increasing the staging rate to 2.44 $\pm$ TB/hour. The difference to the first batch is due to the time that newly activated pools worked.

If enough tapes are requested only the first bottleneck has to be thought about. Mitigating this bottleneck could easily be achieved by increasing the number of tapes and pools. This is, however, not possible in the Jülich LTA due to lack of funding.

The second bottleneck could be mitigated by employing multiple replicas of tapes and using a software that can manage to choose a replica to stage files from if the primary replica is currently in use.

Due to the same reason as for the first bottleneck, this is currently not employed in the Jülich LTA.

\subsection{Reflection on data volumes and computing time}
\label{subsec:Reflection_on_data_volumes_and_computing_time}
The size of the data plays an important role. For some use cases, it might make sense to simply download the data onto the user's personal computer. However, more often, the processing of LOFAR data is an I/O-intensive process, and it can not be performed either on a single powerful machine or on small computing clusters. Instead, high-performance distributed computing systems close to the storage sites are needed (see also Subsection \ref{subsec:Storage_of_final_products}).

\section{Next steps in long-term archiving}
\label{Sect:next}

End of March 2023, Germany officially entered the SKA Observatory (SKAO)\footnote{\url{https://www.skao.int/en/news/459/germany-announces-intention-become-full-ska-observatory-member}}. This means that radio astronomers from Germany can lead Key Science Project proposals and opens the opportunity to host (part of) one of the SKA Regional Centres that will archive SKA data, act as an access point to the data and also further process the data products provided by the Science Data Processor, the high-performance data centres in South Africa and Australia that will pre-process the raw data of the telescopes. The huge amount of data that will be generated in the next years by the SKA means that each year several hundred PB of data need to be archived. This amount of data will be distributed over several Regional Centers, but nevertheless each data centre will need to deal with amounts of data that exceed the current LTA data rates by about a factor of 100.\footnote{\url{https://www.skao.int/en/explore/big-data/362/ska-regional-centres}} Furthermore, this huge amount of data needs to be made accessible to the observers and needs to be analysed.  

Our long years of experience gained from operating the LTA -- JSC started to set up the Jülich LTA in 2009 --  will be extremely valuable in meeting this new data challenge. The current setup is a good starting point to develop all aspects of data storage further. Based on these experiences, we make the following suggestions for future archiving: 

\subsection*{Hardware}
    \begin{itemize}
    \item The hardware needs to handle the enormous amount of data. A trade-off might need to occur between scaling-up the system, such as enhancing the network bandwidth and cache pool performance, and scaling-out, for example by increasing the capacity to retain more data in the cache without the necessity to evict them rapidly after ingestion into the archive. We plan to build a prototype archive that is optimised for storing larger amounts of data than currently the Jülich LTA. This prototype archive will store MeerKAT\footnote{\url{https://www.sarao.ac.za/science/meerkat/about-meerkat/}} \citep{2016mks..confE.....} data. 
    \item As the amount of data increases it is necessary that the data storage is performed in the most energy-efficient way possible. Apart from looking for advanced technical solutions this also implies to develop new data handling strategies.
    \end{itemize}
    
\subsection*{Software}
    \begin{itemize}
    \item With increasing amount of data not only the hardware but also the software needs to be adopted to the larger volume of data. This is not limited to the software used to analyse the data, but also comprises the software related to the (pre-)processing of the data. With ever increasing amount of observational data we will find ourselves soon in a situation where we will no longer be able to store all data, but will need to identify on the fly the data that are worth keeping, similar to what nowadays is already the case at the large particle colliders like e.g.\ LHC. 
    \item A further challenge is that the software to process and analyse the data needs to work in the most energy-efficient way possible. 
    \item One way to achieve these goals would be to develop algorithms that can extract Smart Data\footnote{Data sets that have been extracted from large amounts of data using algorithms according to specific structures and contain meaningful information.} out of the extreme amount of data. Therefore new technologies and methods, including artificial intelligence and machine learning, need to be investigated within the community. These points exemplify the importance of involving the community in the discussion about the characteristics of future archives.
    \item A further aspect that should be considered here, is what is called "Code-to-Data". This buzzword means that the code to reduce and analyse the data should be brought to the data storage sites and be run on computational resources provided there, instead of downloading the data to process them locally. The idea behind this, is that the analysis software and final data products are much smaller in file size than the data sets that are analysed and therefore can be easier transferred, reducing the amount of data traffic and network band width needs (see also example mentioned in Subsection \ref{subsec:Storage_of_final_products}). 
    As already seen, here in Jülich we already provide HPC resources for processing the LOFAR LTA data. While available, optimisation of the workflow of this process is still required. One of the goals of PUNCH4NFDI\footnote{The consortium of particle, astro-, astroparticle, hadron, and nuclear physics within the German National Research Data Infrastructure (Nationale Forschungs-Daten Infrastruktur or NFDI in German; \url{https://www.nfdi.de/?lang=en}); \url{https://www.punch4nfdi.de/}} is to set up a prototype of a federated infrastructure to store and process data remotely. Such a federated infrastructure ("data lake") is also foreseen for the SKAO.
    \item JSC is currently building Europe's first Exascale computer\footnote{\url{https://www.fz-juelich.de/en/ias/jsc/jupiter/tech}}, which has a dynamic Modular System Architecture (dMSA) consisting of two compute modules, a Booster and a Cluster Module. For data applications, the Booster Module is the more relevant one. It will deliver 1 ExaFLOP/s FP64 performance due to its highly scalable system architecture based on the latest generation of GPUs. It will be complemented by a 21 Petabyte Flash Module. At the core of the system, the InfiniBand NDR network connects 25 DragonFly+ groups in the Booster module, as well as 2 extra groups in total for the Cluster module, storage, and administrative infrastructure. In future, additional modules will be added to the system, among them a quantum module and a neuromorphic module. Especially the latter will bring a change from code-to-data to code-with-data approach.\footnote{\url{https://www.fz-juelich.de/en/research/information-and-the-brain/neuromorphic-computing}} 
    \end{itemize}
    
\subsection*{Storage of final data products}
    \begin{itemize}
    \item Future archiving presents several challenges, with one critical aspect being the storage of final data products (see also Subsection \ref{subsec:Storage_of_final_products}). After scientists process their data, it culminates in the creation of a final data product. The long-term archive should evolve by incorporating new data products resulting from data processing, thus adding value to its contents. The flexibility of the dCache software currently allows for the logical and physical separation of data and the ability to treat it differently. It is feasible to establish distinct pools for data originating from various users or producers, each subject to different storage policies (e.g., data stored solely on disk, disk and tape, multiple replicas, etc.). Consequently, while not the primary purpose of the archive, data policies could be reconsidered to accommodate the addition of post-processed products to the archive. This would enhance its value and facilitate scientific research by simplifying access to LOFAR data for future users. For the upcoming LOFAR\,2.0 the LOFAR ERIC foresees to only archive advanced data products indefinitely, while raw and pre-processed data will only be stored for limited amount of time\footnote{\url{https://www.lofar.eu/wp-content/uploads/2023/06/Data_Management_Capabilities.pdf}}. It is also planned to include advanced data products from external processing by science teams into the LTA. Similar considerations are addressed in the design of the SRCNet\footnote{\url{https://aussrc.org/wp-content/uploads/2021/05/SRC-White-Paper-v1.0-Final.pdf}}. Being in the era of multi-wavelengths and multi-messenger astronomy providing simplified  access to data products to colleagues that are not experts in the field of low frequency radio astronomy is an imperative.
    
    \item Additionally, to ensure proper handling of research data such as final data products, clear and enforced guidelines for Research Data Management (RDM) are essential. Key aspects include organisation, documentation, storage, security, data sharing, collaborations, ethical and legal considerations, data preservation, and the implementation of a Data Management Plan (DMP). As an example, Forschungszentrum Jülich published official guidelines in 2019\footnote{\url{https://www.fz-juelich.de/en/zb/open-science/research-data-management}}, which are currently in force and align with the mission of responsibly managing and utilising research data.
    \end{itemize}
    
\subsection*{Providing FAIR data}
    \begin{itemize}
    \item One important goal here is to embrace the FAIR principles \citep{2016NatSD...360018W}, making this final product Findable, Accessible, Interoperable, and Reusable. The significance of adhering to these principles  -- also stated in the LOFAR ERIC Data Policy\footnote{\url{https://www.lofar.eu/wp-content/uploads/2023/06/Data-policy_LOFAR-ERIC.pdf}} and considered in the design of the SRCNet\footnote{\url{https://aussrc.org/wp-content/uploads/2021/05/SRC-White-Paper-v1.0-Final.pdf}} \footnote{\url{http://opensciencefair.eu/images/posters/OSFair2019_paper_102.pdf}}-- lies in their ability to facilitate scientific progress by enabling the reproducibility, replication, and reusability of research outcomes. This, in turn, relies heavily on the long-term accessibility and availability of the underlying data (see \textit{Long-term storage} below). 
    \item To find and reuse data it is indispensable to enrich them with meaningful metadata (see also below). To find a data set the metadata need to provide detailed information about its content. Without well documented observing conditions and reduction steps data sets are of very limited use to other users.
    \item To ensure the interoperability of data an essential prerequisite is the usage of common data formats and standard protocols and to make available standardised software to reduce and analyse data. A data format widely used in astronomy is the Flexible Image Transport System (FITS), which was developed specifically for astronomical data with regard to long-term storage. LOFAR visibility data are stored in the Hierarchical Data Format 5 (HDF5) as it is better suited to store very large data sets.\footnote{\url{https://www.astron.nl/lofarwiki/lib/exe/fetch.php?media=public:documents:lofar-usg-icd-007.pdf}} Defining standards about data (and also metadata) in astronomy is a task of the International Virtual Observatory Alliance (IVOA)\footnote{\url{https://www.ivoa.net/}}. They are also providing software tools that can be used to analyse data of different instruments. In light of "Code-to-Data" one may want to consider to provide software repositories where data reduction and analysis tools are provided that are optimised for the machines on which they run. This will also help to ensure the interoperability of data and facilitate its reuse. However, such an approach requires either that the tools are provided by a central institution or the willingness of the community to share code.
       \end{itemize}
 
\subsection*{Metadata}
    \begin{itemize}
    \item A further aspect involves the correct use of metadata, which enhances the meaning of the data and enables assessment and processing by future users, independent of those who initially created it. The field of observational astronomy, recommendations on the definition and usage of metadata (including metadata standards) are provided by the IVOA and by the Common Archive Observations Model (CAOM)\footnote{\url{https://www.opencadc.org/caom2/}}. For the whole science community, the Helmholtz Metadata Collaboration (HMC)\footnote{\url{https://helmholtz-metadaten.de/en}} develops and implements the enriching of research data by means of metadata adhering to the FAIR principles.
    Within PUNCH4NFDI several task areas are working on topics related to metadata \citep{oelkers_meta,redelbach_2024_10692169}. They investigate for example the role of metadata with respect to data irreversibility or develop concepts for storing metadata in the petabyte range. 
    \end{itemize}

\subsection*{Long-term storage}
    \begin{itemize}
    \item PUNCH4NFDI central aim is not only to store the data but make them FAIR ready. In this context a long-term archive should not merely store data; it should preserve it. Achieving this requires a proper funding model to facilitate preservation and ensure it remains unaffected by political decisions or shifts in strategic direction. Funding should not only sustain the operation of an archive but also anticipate and support the migration of data to new media when the risk of obsolescence threatens data accessibility. 
    \end{itemize}

\subsection*{Monitoring}
    \begin{itemize}
    \item As the system continues to grow in terms of servers and complexity, it becomes necessary to implement additional monitoring mechanisms and improve existing ones. For example, by adding extra dashboards, such as Grafana\footnote{\url{https://grafana.com/}} offers. Certain areas, like energy consumption, have not been fully addressed and require careful monitoring. They are likely to play a crucial role in future implementations. 
    \item Innovative ideas, such as providing users with direct feedback on their energy usage or carbon footprint, are already being discussed. There are even groundbreaking concepts regarding how to allocate compute resources, such as applying for watt-hours instead of core-hours. While this concept is still in development and currently being discussed within computing time allocation, it holds the potential to be expanded and applied to other disciplines, such as the long-term archive. However, there is still work to be done, including the definition of meaningful units.
 
\end{itemize}

\section{Summary and Conclusion}

The Jülich LOFAR Long-term Archive is one of the three data sites that collect and store data for long-term preservation from the Low-Frequency Array (LOFAR) radio telescope. The archive is a complex system consisting of multiple components that interact with each other. The data gathered at the telescope undergoes initial processing at a central facility and is subsequently transmitted to the archive through dedicated network paths. The LTA is a distributed heterogeneous storage system with distinct pools for "hot" and "cold" data. A sophisticated system is in place to enable users to download and work with LOFAR data, for instance, by bringing it online to one of the "hot" data pools.

Built from the ground up, the archive has evolved over the years into its current configuration. Many lessons have been learned through trial and error, and adaptation to new technologies has been crucial to its success. The valuable experience gained will prove essential in addressing the numerous challenges that future telescopes will present, including capacity and production rate issues, as well as unforeseen challenges. Some of the challenges that are beginning to emerge include energy optimisation, which is starting to be addressed through the use of energy-efficient hardware and software. Other challenges, such as bottlenecks, will become more pressing in the next stages of long-term archiving, especially as data rates increase significantly. Securing funding and ensuring data preservation will be pivotal, and new models will need to be developed to address the long-term sustainability of the data.

\section*{Acknowledgement}
We would like to thank our colleagues O.~Mextorf for providing Figure~\ref{fig:JSC_Network} and S.~Graf for his thoughtful comments on an early version of this manuscript which helped us to improve the presentation of its content. Furthermore, we would like to thank the referees for their thoughtful comments that helped us to improve the clarity of our paper. This work was funded by the Deutsche Forschungsgemeinschaft (DFG, German Research Foundation) – project number 460248186 (PUNCH4NFDI) and by the project ``NRW-Cluster for data intensive radio astronomy: Big Bang to Big Data (B$^3$D)'' funded through the programme ``Profilbildung 2020", an initiative of the Ministry of Culture and Science of the State of North Rhine-Westphalia. 

\bibliographystyle{elsarticle-harv} 
\bibliography{section.bib}

\begin{thebibliography}{17}
\expandafter\ifx\csname natexlab\endcsname\relax\def\natexlab#1{#1}\fi
\providecommand{\url}[1]{\texttt{#1}}
\providecommand{\href}[2]{#2}
\providecommand{\path}[1]{#1}
\providecommand{\DOIprefix}{doi:}
\providecommand{\ArXivprefix}{arXiv:}
\providecommand{\URLprefix}{URL: }
\providecommand{\Pubmedprefix}{pmid:}
\providecommand{\doi}[1]{\href{http://dx.doi.org/#1}{\path{#1}}}
\providecommand{\Pubmed}[1]{\href{pmid:#1}{\path{#1}}}
\providecommand{\bibinfo}[2]{#2}
\ifx\xfnm\relax \def\xfnm[#1]{\unskip,\space#1}\fi
\bibitem[{Begeman et~al.(2012)Begeman, Belikov, Boxhoorn, Dijkstra, Holties,
  Renting, Vermaas and Vriend}]{2012Begeman_Astro-WISE}
\bibinfo{author}{Begeman, K.}, \bibinfo{author}{Belikov, A.},
  \bibinfo{author}{Boxhoorn, D.}, \bibinfo{author}{Dijkstra, F.},
  \bibinfo{author}{Holties, H.}, \bibinfo{author}{Renting, G.},
  \bibinfo{author}{Vermaas, N.}, \bibinfo{author}{Vriend, W.},
  \bibinfo{year}{2012}.
\newblock \bibinfo{title}{Scaling astro-wise to lofar long term archive}.
\newblock \bibinfo{journal}{Experimental Astronomy} \bibinfo{volume}{35}.
\newblock \DOIprefix\doi{10.1007/s10686-012-9305-2}.
\bibitem[{Deutsch and Gailly(1996)}]{rfc1950}
\bibinfo{author}{Deutsch, L.P.}, \bibinfo{author}{Gailly, J.L.},
  \bibinfo{year}{1996}.
\newblock \bibinfo{title}{{ZLIB Compressed Data Format Specification version
  3.3}}.
\newblock \bibinfo{howpublished}{RFC 1950}.
\newblock \URLprefix \url{https://www.rfc-editor.org/info/rfc1950},
  \DOIprefix\doi{10.17487/RFC1950}.
\bibitem[{Drabent et~al.(2020)Drabent, Hoeft, Mechev, Oonk, Shimwell, Sweijen,
  Danezi, Schrijvers, Manzano, Tsigenov, Dettmar, Brüggen and
  Schwarz}]{Drabent:874406}
\bibinfo{author}{Drabent, A.}, \bibinfo{author}{Hoeft, M.},
  \bibinfo{author}{Mechev, A.B.}, \bibinfo{author}{Oonk, J.B.R.},
  \bibinfo{author}{Shimwell, T.W.}, \bibinfo{author}{Sweijen, F.},
  \bibinfo{author}{Danezi, A.}, \bibinfo{author}{Schrijvers, C.},
  \bibinfo{author}{Manzano, C.}, \bibinfo{author}{Tsigenov, O.},
  \bibinfo{author}{Dettmar, R.J.}, \bibinfo{author}{Brüggen, M.},
  \bibinfo{author}{Schwarz, D.J.}, \bibinfo{year}{2020}.
\newblock \bibinfo{title}{{R}ealising the {LOFAR} {T}wo-{M}etre {S}ky
  {S}urvey}, in: \bibinfo{booktitle}{NIC Symposium 2020},
  \bibinfo{organization}{NIC Symposium 2020, Jülich (Germany), 27 Feb 2020 -
  28 Feb 2020}. \bibinfo{publisher}{Forschungszentrum Jülich GmbH
  Zentralbibliothek, Verlag}, \bibinfo{address}{Jülich}. pp.
  \bibinfo{pages}{113 -- 122}.
\newblock \URLprefix \url{https://juser.fz-juelich.de/record/874406}.
\bibitem[{Fletcher(1982)}]{DBLP:journals/tcom/Fletcher82}
\bibinfo{author}{Fletcher, J.G.}, \bibinfo{year}{1982}.
\newblock \bibinfo{title}{An arithmetic checksum for serial transmissions}.
\newblock \bibinfo{journal}{{IEEE} Trans. Commun.} \bibinfo{volume}{30},
  \bibinfo{pages}{247--252}.
\newblock \URLprefix \url{https://doi.org/10.1109/TCOM.1982.1095369},
  \DOIprefix\doi{10.1109/TCOM.1982.1095369}.
\bibitem[{{Inside Big Data} and {Device Apps SMART
  Cities}(2017)}]{data2017exponential}
\bibinfo{author}{{Inside Big Data}}, \bibinfo{author}{{Device Apps SMART
  Cities}}, \bibinfo{year}{2017}.
\newblock \bibinfo{title}{The exponential growth of data}.
\newblock \bibinfo{journal}{Inside Big Data White paper.} \URLprefix
  \url{https://insidebigdata.com/2017/02/16/the-exponential-growth-of-data}.
\bibitem[{{Jonas} and {MeerKAT Team}(2016)}]{2016mks..confE.....}
\bibinfo{author}{{Jonas}, J.}, \bibinfo{author}{{MeerKAT Team}},
  \bibinfo{year}{2016}.
\newblock \bibinfo{title}{{The MeerKAT radio telescope}}, in:
  \bibinfo{booktitle}{{MeerKAT Science: On the Pathway to the SKA}}.
\newblock \URLprefix \url{https://pos.sissa.it/277/001/pdf}.
\bibitem[{{Oelkers}(2023)}]{oelkers_meta}
\bibinfo{author}{{Oelkers}, T.}, \bibinfo{year}{2023}.
\newblock \bibinfo{title}{Overview of petabyte-scale metadata storage methods
  and frameworks}.
\newblock \bibinfo{journal}{PUNCH4NFDI report} \URLprefix
  \url{https://results.punch4nfdi.de/files/documents/Metadata_PUNCH_Oelkers.pdf}.
\bibitem[{{Offringa}(2016)}]{2016A&A...595A..99O}
\bibinfo{author}{{Offringa}, A.R.}, \bibinfo{year}{2016}.
\newblock \bibinfo{title}{{Compression of interferometric radio-astronomical
  data}}.
\newblock \bibinfo{journal}{\aap} \bibinfo{volume}{595}, \bibinfo{pages}{A99}.
\newblock \DOIprefix\doi{10.1051/0004-6361/201629565},
  \href{http://arxiv.org/abs/1609.02019}{{\tt arXiv:1609.02019}}.
\bibitem[{Redelbach et~al.(2024)Redelbach, Dembinski, Hessling, Karuppusamy,
  Kramer, Lenok, Nordin, Pfalzner, Schwarz, Straessner and
  Vybornov}]{redelbach_2024_10692169}
\bibinfo{author}{Redelbach, A.}, \bibinfo{author}{Dembinski, H.},
  \bibinfo{author}{Hessling, H.}, \bibinfo{author}{Karuppusamy, R.},
  \bibinfo{author}{Kramer, M.}, \bibinfo{author}{Lenok, V.},
  \bibinfo{author}{Nordin, J.}, \bibinfo{author}{Pfalzner, S.},
  \bibinfo{author}{Schwarz, D.}, \bibinfo{author}{Straessner, A.},
  \bibinfo{author}{Vybornov, V.}, \bibinfo{year}{2024}.
\newblock \bibinfo{title}{{Curation and metadata - concepts for data
  irreversibility}}.
\newblock \URLprefix \url{https://doi.org/10.5281/zenodo.10692169},
  \DOIprefix\doi{10.5281/zenodo.10692169}.
\bibitem[{{Renting} and {Holties}(2011)}]{2011ASPC..442...49R}
\bibinfo{author}{{Renting}, G.A.}, \bibinfo{author}{{Holties}, H.A.},
  \bibinfo{year}{2011}.
\newblock \bibinfo{title}{{LOFAR Long Term Archive}}, in:
  \bibinfo{editor}{{Evans}, I.N.}, \bibinfo{editor}{{Accomazzi}, A.},
  \bibinfo{editor}{{Mink}, D.J.}, \bibinfo{editor}{{Rots}, A.H.} (Eds.),
  \bibinfo{booktitle}{Astronomical Data Analysis Software and Systems XX},
  p.~\bibinfo{pages}{49}.
\bibitem[{Rivest(1992)}]{rfc1321}
\bibinfo{author}{Rivest, R.L.}, \bibinfo{year}{1992}.
\newblock \bibinfo{title}{{The MD5 Message-Digest Algorithm}}.
\newblock \bibinfo{howpublished}{RFC 1321}.
\newblock \URLprefix \url{https://www.rfc-editor.org/info/rfc1321},
  \DOIprefix\doi{10.17487/RFC1321}.
\bibitem[{{Shimwell} et~al.(2019){Shimwell}, {Tasse}, {Hardcastle}, {Mechev},
  {Williams}, {Best}, {R{\"o}ttgering}, {Callingham}, {Dijkema}, {de Gasperin},
  {Hoang}, {Hugo}, {Mirmont}, {Oonk}, {Prandoni}, {Rafferty}, {Sabater},
  {Smirnov}, {van Weeren}, {White}, {Atemkeng}, {Bester}, {Bonnassieux},
  {Br{\"u}ggen}, {Brunetti}, {Chy{\.z}y}, {Cochrane}, {Conway}, {Croston},
  {Danezi}, {Duncan}, {Haverkorn}, {Heald}, {Iacobelli}, {Intema}, {Jackson},
  {Jamrozy}, {Jarvis}, {Lakhoo}, {Mevius}, {Miley}, {Morabito}, {Morganti},
  {Nisbet}, {Orr{\'u}}, {Perkins}, {Pizzo}, {Schrijvers}, {Smith}, {Vermeulen},
  {Wise}, {Alegre}, {Bacon}, {van Bemmel}, {Beswick}, {Bonafede}, {Botteon},
  {Bourke}, {Brienza}, {Calistro Rivera}, {Cassano}, {Clarke}, {Conselice},
  {Dettmar}, {Drabent}, {Dumba}, {Emig}, {En{\ss}lin}, {Ferrari}, {Garrett},
  {G{\'e}nova-Santos}, {Goyal}, {G{\"u}rkan}, {Hale}, {Harwood}, {Heesen},
  {Hoeft}, {Horellou}, {Jackson}, {Kokotanekov}, {Kondapally},
  {Kunert-Bajraszewska}, {Mahatma}, {Mahony}, {Mandal}, {McKean}, {Merloni},
  {Mingo}, {Miskolczi}, {Mooney}, {Nikiel-Wroczy{\'n}ski}, {O'Sullivan},
  {Quinn}, {Reich}, {Roskowi{\'n}ski}, {Rowlinson}, {Savini}, {Saxena},
  {Schwarz}, {Shulevski}, {Sridhar}, {Stacey}, {Urquhart}, {van der Wiel},
  {Varenius}, {Webster} and {Wilber}}]{2019A&A...622A...1S}
\bibinfo{author}{{Shimwell}, T.W.}, \bibinfo{author}{{Tasse}, C.},
  \bibinfo{author}{{Hardcastle}, M.J.}, \bibinfo{author}{{Mechev}, A.P.},
  \bibinfo{author}{{Williams}, W.L.}, \bibinfo{author}{{Best}, P.N.},
  \bibinfo{author}{{R{\"o}ttgering}, H.J.A.}, \bibinfo{author}{{Callingham},
  J.R.}, \bibinfo{author}{{Dijkema}, T.J.}, \bibinfo{author}{{de Gasperin},
  F.}, \bibinfo{author}{{Hoang}, D.N.}, \bibinfo{author}{{Hugo}, B.},
  \bibinfo{author}{{Mirmont}, M.}, \bibinfo{author}{{Oonk}, J.B.R.},
  \bibinfo{author}{{Prandoni}, I.}, \bibinfo{author}{{Rafferty}, D.},
  \bibinfo{author}{{Sabater}, J.}, \bibinfo{author}{{Smirnov}, O.},
  \bibinfo{author}{{van Weeren}, R.J.}, \bibinfo{author}{{White}, G.J.},
  \bibinfo{author}{{Atemkeng}, M.}, \bibinfo{author}{{Bester}, L.},
  \bibinfo{author}{{Bonnassieux}, E.}, \bibinfo{author}{{Br{\"u}ggen}, M.},
  \bibinfo{author}{{Brunetti}, G.}, \bibinfo{author}{{Chy{\.z}y}, K.T.},
  \bibinfo{author}{{Cochrane}, R.}, \bibinfo{author}{{Conway}, J.E.},
  \bibinfo{author}{{Croston}, J.H.}, \bibinfo{author}{{Danezi}, A.},
  \bibinfo{author}{{Duncan}, K.}, \bibinfo{author}{{Haverkorn}, M.},
  \bibinfo{author}{{Heald}, G.H.}, \bibinfo{author}{{Iacobelli}, M.},
  \bibinfo{author}{{Intema}, H.T.}, \bibinfo{author}{{Jackson}, N.},
  \bibinfo{author}{{Jamrozy}, M.}, \bibinfo{author}{{Jarvis}, M.J.},
  \bibinfo{author}{{Lakhoo}, R.}, \bibinfo{author}{{Mevius}, M.},
  \bibinfo{author}{{Miley}, G.K.}, \bibinfo{author}{{Morabito}, L.},
  \bibinfo{author}{{Morganti}, R.}, \bibinfo{author}{{Nisbet}, D.},
  \bibinfo{author}{{Orr{\'u}}, E.}, \bibinfo{author}{{Perkins}, S.},
  \bibinfo{author}{{Pizzo}, R.F.}, \bibinfo{author}{{Schrijvers}, C.},
  \bibinfo{author}{{Smith}, D.J.B.}, \bibinfo{author}{{Vermeulen}, R.},
  \bibinfo{author}{{Wise}, M.W.}, \bibinfo{author}{{Alegre}, L.},
  \bibinfo{author}{{Bacon}, D.J.}, \bibinfo{author}{{van Bemmel}, I.M.},
  \bibinfo{author}{{Beswick}, R.J.}, \bibinfo{author}{{Bonafede}, A.},
  \bibinfo{author}{{Botteon}, A.}, \bibinfo{author}{{Bourke}, S.},
  \bibinfo{author}{{Brienza}, M.}, \bibinfo{author}{{Calistro Rivera}, G.},
  \bibinfo{author}{{Cassano}, R.}, \bibinfo{author}{{Clarke}, A.O.},
  \bibinfo{author}{{Conselice}, C.J.}, \bibinfo{author}{{Dettmar}, R.J.},
  \bibinfo{author}{{Drabent}, A.}, \bibinfo{author}{{Dumba}, C.},
  \bibinfo{author}{{Emig}, K.L.}, \bibinfo{author}{{En{\ss}lin}, T.A.},
  \bibinfo{author}{{Ferrari}, C.}, \bibinfo{author}{{Garrett}, M.A.},
  \bibinfo{author}{{G{\'e}nova-Santos}, R.T.}, \bibinfo{author}{{Goyal}, A.},
  \bibinfo{author}{{G{\"u}rkan}, G.}, \bibinfo{author}{{Hale}, C.},
  \bibinfo{author}{{Harwood}, J.J.}, \bibinfo{author}{{Heesen}, V.},
  \bibinfo{author}{{Hoeft}, M.}, \bibinfo{author}{{Horellou}, C.},
  \bibinfo{author}{{Jackson}, C.}, \bibinfo{author}{{Kokotanekov}, G.},
  \bibinfo{author}{{Kondapally}, R.}, \bibinfo{author}{{Kunert-Bajraszewska},
  M.}, \bibinfo{author}{{Mahatma}, V.}, \bibinfo{author}{{Mahony}, E.K.},
  \bibinfo{author}{{Mandal}, S.}, \bibinfo{author}{{McKean}, J.P.},
  \bibinfo{author}{{Merloni}, A.}, \bibinfo{author}{{Mingo}, B.},
  \bibinfo{author}{{Miskolczi}, A.}, \bibinfo{author}{{Mooney}, S.},
  \bibinfo{author}{{Nikiel-Wroczy{\'n}ski}, B.}, \bibinfo{author}{{O'Sullivan},
  S.P.}, \bibinfo{author}{{Quinn}, J.}, \bibinfo{author}{{Reich}, W.},
  \bibinfo{author}{{Roskowi{\'n}ski}, C.}, \bibinfo{author}{{Rowlinson}, A.},
  \bibinfo{author}{{Savini}, F.}, \bibinfo{author}{{Saxena}, A.},
  \bibinfo{author}{{Schwarz}, D.J.}, \bibinfo{author}{{Shulevski}, A.},
  \bibinfo{author}{{Sridhar}, S.S.}, \bibinfo{author}{{Stacey}, H.R.},
  \bibinfo{author}{{Urquhart}, S.}, \bibinfo{author}{{van der Wiel}, M.H.D.},
  \bibinfo{author}{{Varenius}, E.}, \bibinfo{author}{{Webster}, B.},
  \bibinfo{author}{{Wilber}, A.}, \bibinfo{year}{2019}.
\newblock \bibinfo{title}{{The LOFAR Two-metre Sky Survey. II. First data
  release}}.
\newblock \bibinfo{journal}{\aap} \bibinfo{volume}{622}, \bibinfo{pages}{A1}.
\newblock \DOIprefix\doi{10.1051/0004-6361/201833559},
  \href{http://arxiv.org/abs/1811.07926}{{\tt arXiv:1811.07926}}.
\bibitem[{{Szalay} and {Gray}(2006)}]{2006Natur.440..413S}
\bibinfo{author}{{Szalay}, A.}, \bibinfo{author}{{Gray}, J.},
  \bibinfo{year}{2006}.
\newblock \bibinfo{title}{{2020 Computing: Science in an exponential world}}.
\newblock \bibinfo{journal}{\nat} \bibinfo{volume}{440},
  \bibinfo{pages}{413--414}.
\newblock \DOIprefix\doi{10.1038/440413a}.
\bibitem[{{Valentijn} and {Belikov}(2009)}]{2009MmSAI..80..509V}
\bibinfo{author}{{Valentijn}, E.}, \bibinfo{author}{{Belikov}, A.N.},
  \bibinfo{year}{2009}.
\newblock \bibinfo{title}{{Lofar information system design .}}
\newblock \bibinfo{journal}{\memsai} \bibinfo{volume}{80},
  \bibinfo{pages}{509}.
\bibitem[{Van~Cappellen et~al.(2023)Van~Cappellen, Saavedra, Hut, Gunst,
  Schoenmakers and Veen}]{10265351}
\bibinfo{author}{Van~Cappellen, W.A.}, \bibinfo{author}{Saavedra, C.J.B.},
  \bibinfo{author}{Hut, B.}, \bibinfo{author}{Gunst, A.W.},
  \bibinfo{author}{Schoenmakers, A.P.}, \bibinfo{author}{Veen, S.T.},
  \bibinfo{year}{2023}.
\newblock \bibinfo{title}{Innovations in lofar: An overview of improvements of
  the lofar telescope}, in: \bibinfo{booktitle}{2023 XXXVth General Assembly
  and Scientific Symposium of the International Union of Radio Science (URSI
  GASS)}, pp. \bibinfo{pages}{1--3}.
\newblock \DOIprefix\doi{10.23919/URSIGASS57860.2023.10265351}.
\bibitem[{{van Haarlem} et~al.(2013){van Haarlem}, {Wise}, {Gunst}, {Heald},
  {McKean}, {Hessels}, {de Bruyn}, {Nijboer}, {Swinbank}, {Fallows},
  {Brentjens}, {Nelles}, {Beck}, {Falcke}, {Fender}, {H{\"o}randel},
  {Koopmans}, {Mann}, {Miley}, {R{\"o}ttgering}, {Stappers}, {Wijers},
  {Zaroubi}, {van den Akker}, {Alexov}, {Anderson}, {Anderson}, {van Ardenne},
  {Arts}, {Asgekar}, {Avruch}, {Batejat}, {B{\"a}hren}, {Bell}, {Bell}, {van
  Bemmel}, {Bennema}, {Bentum}, {Bernardi}, {Best}, {B{\^\i}rzan}, {Bonafede},
  {Boonstra}, {Braun}, {Bregman}, {Breitling}, {van de Brink}, {Broderick},
  {Broekema}, {Brouw}, {Br{\"u}ggen}, {Butcher}, {van Cappellen}, {Ciardi},
  {Coenen}, {Conway}, {Coolen}, {Corstanje}, {Damstra}, {Davies}, {Deller},
  {Dettmar}, {van Diepen}, {Dijkstra}, {Donker}, {Doorduin}, {Dromer}, {Drost},
  {van Duin}, {Eisl{\"o}ffel}, {van Enst}, {Ferrari}, {Frieswijk}, {Gankema},
  {Garrett}, {de Gasperin}, {Gerbers}, {de Geus}, {Grie{\ss}meier}, {Grit},
  {Gruppen}, {Hamaker}, {Hassall}, {Hoeft}, {Holties}, {Horneffer}, {van der
  Horst}, {van Houwelingen}, {Huijgen}, {Iacobelli}, {Intema}, {Jackson},
  {Jelic}, {de Jong}, {Juette}, {Kant}, {Karastergiou}, {Koers}, {Kollen},
  {Kondratiev}, {Kooistra}, {Koopman}, {Koster}, {Kuniyoshi}, {Kramer},
  {Kuper}, {Lambropoulos}, {Law}, {van Leeuwen}, {Lemaitre}, {Loose}, {Maat},
  {Macario}, {Markoff}, {Masters}, {McFadden}, {McKay-Bukowski}, {Meijering},
  {Meulman}, {Mevius}, {Middelberg}, {Millenaar}, {Miller-Jones}, {Mohan},
  {Mol}, {Morawietz}, {Morganti}, {Mulcahy}, {Mulder}, {Munk}, {Nieuwenhuis},
  {van Nieuwpoort}, {Noordam}, {Norden}, {Noutsos}, {Offringa}, {Olofsson},
  {Omar}, {Orr{\'u}}, {Overeem}, {Paas}, {Pandey-Pommier}, {Pandey}, {Pizzo},
  {Polatidis}, {Rafferty}, {Rawlings}, {Reich}, {de Reijer}, {Reitsma},
  {Renting}, {Riemers}, {Rol}, {Romein}, {Roosjen}, {Ruiter}, {Scaife}, {van
  der Schaaf}, {Scheers}, {Schellart}, {Schoenmakers}, {Schoonderbeek},
  {Serylak}, {Shulevski}, {Sluman}, {Smirnov}, {Sobey}, {Spreeuw}, {Steinmetz},
  {Sterks}, {Stiepel}, {Stuurwold}, {Tagger}, {Tang}, {Tasse}, {Thomas},
  {Thoudam}, {Toribio}, {van der Tol}, {Usov}, {van Veelen}, {van der Veen},
  {ter Veen}, {Verbiest}, {Vermeulen}, {Vermaas}, {Vocks}, {Vogt}, {de Vos},
  {van der Wal}, {van Weeren}, {Weggemans}, {Weltevrede}, {White}, {Wijnholds},
  {Wilhelmsson}, {Wucknitz}, {Yatawatta}, {Zarka}, {Zensus} and {van
  Zwieten}}]{2013A&A...556A...2V}
\bibinfo{author}{{van Haarlem}, M.P.}, \bibinfo{author}{{Wise}, M.W.},
  \bibinfo{author}{{Gunst}, A.W.}, \bibinfo{author}{{Heald}, G.},
  \bibinfo{author}{{McKean}, J.P.}, \bibinfo{author}{{Hessels}, J.W.T.},
  \bibinfo{author}{{de Bruyn}, A.G.}, \bibinfo{author}{{Nijboer}, R.},
  \bibinfo{author}{{Swinbank}, J.}, \bibinfo{author}{{Fallows}, R.},
  \bibinfo{author}{{Brentjens}, M.}, \bibinfo{author}{{Nelles}, A.},
  \bibinfo{author}{{Beck}, R.}, \bibinfo{author}{{Falcke}, H.},
  \bibinfo{author}{{Fender}, R.}, \bibinfo{author}{{H{\"o}randel}, J.},
  \bibinfo{author}{{Koopmans}, L.V.E.}, \bibinfo{author}{{Mann}, G.},
  \bibinfo{author}{{Miley}, G.}, \bibinfo{author}{{R{\"o}ttgering}, H.},
  \bibinfo{author}{{Stappers}, B.W.}, \bibinfo{author}{{Wijers}, R.A.M.J.},
  \bibinfo{author}{{Zaroubi}, S.}, \bibinfo{author}{{van den Akker}, M.},
  \bibinfo{author}{{Alexov}, A.}, \bibinfo{author}{{Anderson}, J.},
  \bibinfo{author}{{Anderson}, K.}, \bibinfo{author}{{van Ardenne}, A.},
  \bibinfo{author}{{Arts}, M.}, \bibinfo{author}{{Asgekar}, A.},
  \bibinfo{author}{{Avruch}, I.M.}, \bibinfo{author}{{Batejat}, F.},
  \bibinfo{author}{{B{\"a}hren}, L.}, \bibinfo{author}{{Bell}, M.E.},
  \bibinfo{author}{{Bell}, M.R.}, \bibinfo{author}{{van Bemmel}, I.},
  \bibinfo{author}{{Bennema}, P.}, \bibinfo{author}{{Bentum}, M.J.},
  \bibinfo{author}{{Bernardi}, G.}, \bibinfo{author}{{Best}, P.},
  \bibinfo{author}{{B{\^\i}rzan}, L.}, \bibinfo{author}{{Bonafede}, A.},
  \bibinfo{author}{{Boonstra}, A.J.}, \bibinfo{author}{{Braun}, R.},
  \bibinfo{author}{{Bregman}, J.}, \bibinfo{author}{{Breitling}, F.},
  \bibinfo{author}{{van de Brink}, R.H.}, \bibinfo{author}{{Broderick}, J.},
  \bibinfo{author}{{Broekema}, P.C.}, \bibinfo{author}{{Brouw}, W.N.},
  \bibinfo{author}{{Br{\"u}ggen}, M.}, \bibinfo{author}{{Butcher}, H.R.},
  \bibinfo{author}{{van Cappellen}, W.}, \bibinfo{author}{{Ciardi}, B.},
  \bibinfo{author}{{Coenen}, T.}, \bibinfo{author}{{Conway}, J.},
  \bibinfo{author}{{Coolen}, A.}, \bibinfo{author}{{Corstanje}, A.},
  \bibinfo{author}{{Damstra}, S.}, \bibinfo{author}{{Davies}, O.},
  \bibinfo{author}{{Deller}, A.T.}, \bibinfo{author}{{Dettmar}, R.J.},
  \bibinfo{author}{{van Diepen}, G.}, \bibinfo{author}{{Dijkstra}, K.},
  \bibinfo{author}{{Donker}, P.}, \bibinfo{author}{{Doorduin}, A.},
  \bibinfo{author}{{Dromer}, J.}, \bibinfo{author}{{Drost}, M.},
  \bibinfo{author}{{van Duin}, A.}, \bibinfo{author}{{Eisl{\"o}ffel}, J.},
  \bibinfo{author}{{van Enst}, J.}, \bibinfo{author}{{Ferrari}, C.},
  \bibinfo{author}{{Frieswijk}, W.}, \bibinfo{author}{{Gankema}, H.},
  \bibinfo{author}{{Garrett}, M.A.}, \bibinfo{author}{{de Gasperin}, F.},
  \bibinfo{author}{{Gerbers}, M.}, \bibinfo{author}{{de Geus}, E.},
  \bibinfo{author}{{Grie{\ss}meier}, J.M.}, \bibinfo{author}{{Grit}, T.},
  \bibinfo{author}{{Gruppen}, P.}, \bibinfo{author}{{Hamaker}, J.P.},
  \bibinfo{author}{{Hassall}, T.}, \bibinfo{author}{{Hoeft}, M.},
  \bibinfo{author}{{Holties}, H.A.}, \bibinfo{author}{{Horneffer}, A.},
  \bibinfo{author}{{van der Horst}, A.}, \bibinfo{author}{{van Houwelingen},
  A.}, \bibinfo{author}{{Huijgen}, A.}, \bibinfo{author}{{Iacobelli}, M.},
  \bibinfo{author}{{Intema}, H.}, \bibinfo{author}{{Jackson}, N.},
  \bibinfo{author}{{Jelic}, V.}, \bibinfo{author}{{de Jong}, A.},
  \bibinfo{author}{{Juette}, E.}, \bibinfo{author}{{Kant}, D.},
  \bibinfo{author}{{Karastergiou}, A.}, \bibinfo{author}{{Koers}, A.},
  \bibinfo{author}{{Kollen}, H.}, \bibinfo{author}{{Kondratiev}, V.I.},
  \bibinfo{author}{{Kooistra}, E.}, \bibinfo{author}{{Koopman}, Y.},
  \bibinfo{author}{{Koster}, A.}, \bibinfo{author}{{Kuniyoshi}, M.},
  \bibinfo{author}{{Kramer}, M.}, \bibinfo{author}{{Kuper}, G.},
  \bibinfo{author}{{Lambropoulos}, P.}, \bibinfo{author}{{Law}, C.},
  \bibinfo{author}{{van Leeuwen}, J.}, \bibinfo{author}{{Lemaitre}, J.},
  \bibinfo{author}{{Loose}, M.}, \bibinfo{author}{{Maat}, P.},
  \bibinfo{author}{{Macario}, G.}, \bibinfo{author}{{Markoff}, S.},
  \bibinfo{author}{{Masters}, J.}, \bibinfo{author}{{McFadden}, R.A.},
  \bibinfo{author}{{McKay-Bukowski}, D.}, \bibinfo{author}{{Meijering}, H.},
  \bibinfo{author}{{Meulman}, H.}, \bibinfo{author}{{Mevius}, M.},
  \bibinfo{author}{{Middelberg}, E.}, \bibinfo{author}{{Millenaar}, R.},
  \bibinfo{author}{{Miller-Jones}, J.C.A.}, \bibinfo{author}{{Mohan}, R.N.},
  \bibinfo{author}{{Mol}, J.D.}, \bibinfo{author}{{Morawietz}, J.},
  \bibinfo{author}{{Morganti}, R.}, \bibinfo{author}{{Mulcahy}, D.D.},
  \bibinfo{author}{{Mulder}, E.}, \bibinfo{author}{{Munk}, H.},
  \bibinfo{author}{{Nieuwenhuis}, L.}, \bibinfo{author}{{van Nieuwpoort}, R.},
  \bibinfo{author}{{Noordam}, J.E.}, \bibinfo{author}{{Norden}, M.},
  \bibinfo{author}{{Noutsos}, A.}, \bibinfo{author}{{Offringa}, A.R.},
  \bibinfo{author}{{Olofsson}, H.}, \bibinfo{author}{{Omar}, A.},
  \bibinfo{author}{{Orr{\'u}}, E.}, \bibinfo{author}{{Overeem}, R.},
  \bibinfo{author}{{Paas}, H.}, \bibinfo{author}{{Pandey-Pommier}, M.},
  \bibinfo{author}{{Pandey}, V.N.}, \bibinfo{author}{{Pizzo}, R.},
  \bibinfo{author}{{Polatidis}, A.}, \bibinfo{author}{{Rafferty}, D.},
  \bibinfo{author}{{Rawlings}, S.}, \bibinfo{author}{{Reich}, W.},
  \bibinfo{author}{{de Reijer}, J.P.}, \bibinfo{author}{{Reitsma}, J.},
  \bibinfo{author}{{Renting}, G.A.}, \bibinfo{author}{{Riemers}, P.},
  \bibinfo{author}{{Rol}, E.}, \bibinfo{author}{{Romein}, J.W.},
  \bibinfo{author}{{Roosjen}, J.}, \bibinfo{author}{{Ruiter}, M.},
  \bibinfo{author}{{Scaife}, A.}, \bibinfo{author}{{van der Schaaf}, K.},
  \bibinfo{author}{{Scheers}, B.}, \bibinfo{author}{{Schellart}, P.},
  \bibinfo{author}{{Schoenmakers}, A.}, \bibinfo{author}{{Schoonderbeek}, G.},
  \bibinfo{author}{{Serylak}, M.}, \bibinfo{author}{{Shulevski}, A.},
  \bibinfo{author}{{Sluman}, J.}, \bibinfo{author}{{Smirnov}, O.},
  \bibinfo{author}{{Sobey}, C.}, \bibinfo{author}{{Spreeuw}, H.},
  \bibinfo{author}{{Steinmetz}, M.}, \bibinfo{author}{{Sterks}, C.G.M.},
  \bibinfo{author}{{Stiepel}, H.J.}, \bibinfo{author}{{Stuurwold}, K.},
  \bibinfo{author}{{Tagger}, M.}, \bibinfo{author}{{Tang}, Y.},
  \bibinfo{author}{{Tasse}, C.}, \bibinfo{author}{{Thomas}, I.},
  \bibinfo{author}{{Thoudam}, S.}, \bibinfo{author}{{Toribio}, M.C.},
  \bibinfo{author}{{van der Tol}, B.}, \bibinfo{author}{{Usov}, O.},
  \bibinfo{author}{{van Veelen}, M.}, \bibinfo{author}{{van der Veen}, A.J.},
  \bibinfo{author}{{ter Veen}, S.}, \bibinfo{author}{{Verbiest}, J.P.W.},
  \bibinfo{author}{{Vermeulen}, R.}, \bibinfo{author}{{Vermaas}, N.},
  \bibinfo{author}{{Vocks}, C.}, \bibinfo{author}{{Vogt}, C.},
  \bibinfo{author}{{de Vos}, M.}, \bibinfo{author}{{van der Wal}, E.},
  \bibinfo{author}{{van Weeren}, R.}, \bibinfo{author}{{Weggemans}, H.},
  \bibinfo{author}{{Weltevrede}, P.}, \bibinfo{author}{{White}, S.},
  \bibinfo{author}{{Wijnholds}, S.J.}, \bibinfo{author}{{Wilhelmsson}, T.},
  \bibinfo{author}{{Wucknitz}, O.}, \bibinfo{author}{{Yatawatta}, S.},
  \bibinfo{author}{{Zarka}, P.}, \bibinfo{author}{{Zensus}, A.},
  \bibinfo{author}{{van Zwieten}, J.}, \bibinfo{year}{2013}.
\newblock \bibinfo{title}{{LOFAR: The LOw-Frequency ARray}}.
\newblock \bibinfo{journal}{\aap} \bibinfo{volume}{556}, \bibinfo{pages}{A2}.
\newblock \DOIprefix\doi{10.1051/0004-6361/201220873},
  \href{http://arxiv.org/abs/1305.3550}{{\tt arXiv:1305.3550}}.
\bibitem[{{Wilkinson} et~al.(2016){Wilkinson}, {Dumontier}, {Aalbersberg},
  {Appleton}, {Axton}, {Baak}, {Blomberg}, {Boiten}, {da Silva Santos},
  {Bourne}, {Bouwman}, {Brookes}, {Clark}, {Crosas}, {Dillo}, {Dumon},
  {Edmunds}, {Evelo}, {Finkers}, {Gonzalez-Beltran}, {Gray}, {Groth}, {Goble},
  {Grethe}, {Heringa}, {'T Hoen}, {Hooft}, {Kuhn}, {Kok}, {Kok}, {Lusher},
  {Martone}, {Mons}, {Packer}, {Persson}, {Rocca-Serra}, {Roos}, {van Schaik},
  {Sansone}, {Schultes}, {Sengstag}, {Slater}, {Strawn}, {Swertz}, {Thompson},
  {van der Lei}, {van Mulligen}, {Velterop}, {Waagmeester}, {Wittenburg},
  {Wolstencroft}, {Zhao} and {Mons}}]{2016NatSD...360018W}
\bibinfo{author}{{Wilkinson}, M.D.}, \bibinfo{author}{{Dumontier}, M.},
  \bibinfo{author}{{Aalbersberg}, I.J.}, \bibinfo{author}{{Appleton}, G.},
  \bibinfo{author}{{Axton}, M.}, \bibinfo{author}{{Baak}, A.},
  \bibinfo{author}{{Blomberg}, N.}, \bibinfo{author}{{Boiten}, J.W.},
  \bibinfo{author}{{da Silva Santos}, L.B.}, \bibinfo{author}{{Bourne}, P.E.},
  \bibinfo{author}{{Bouwman}, J.}, \bibinfo{author}{{Brookes}, A.J.},
  \bibinfo{author}{{Clark}, T.}, \bibinfo{author}{{Crosas}, M.},
  \bibinfo{author}{{Dillo}, I.}, \bibinfo{author}{{Dumon}, O.},
  \bibinfo{author}{{Edmunds}, S.}, \bibinfo{author}{{Evelo}, C.T.},
  \bibinfo{author}{{Finkers}, R.}, \bibinfo{author}{{Gonzalez-Beltran}, A.},
  \bibinfo{author}{{Gray}, A.J.G.}, \bibinfo{author}{{Groth}, P.},
  \bibinfo{author}{{Goble}, C.}, \bibinfo{author}{{Grethe}, J.S.},
  \bibinfo{author}{{Heringa}, J.}, \bibinfo{author}{{'T Hoen}, P.A.C.},
  \bibinfo{author}{{Hooft}, R.}, \bibinfo{author}{{Kuhn}, T.},
  \bibinfo{author}{{Kok}, R.}, \bibinfo{author}{{Kok}, J.},
  \bibinfo{author}{{Lusher}, S.J.}, \bibinfo{author}{{Martone}, M.E.},
  \bibinfo{author}{{Mons}, A.}, \bibinfo{author}{{Packer}, A.L.},
  \bibinfo{author}{{Persson}, B.}, \bibinfo{author}{{Rocca-Serra}, P.},
  \bibinfo{author}{{Roos}, M.}, \bibinfo{author}{{van Schaik}, R.},
  \bibinfo{author}{{Sansone}, S.A.}, \bibinfo{author}{{Schultes}, E.},
  \bibinfo{author}{{Sengstag}, T.}, \bibinfo{author}{{Slater}, T.},
  \bibinfo{author}{{Strawn}, G.}, \bibinfo{author}{{Swertz}, M.A.},
  \bibinfo{author}{{Thompson}, M.}, \bibinfo{author}{{van der Lei}, J.},
  \bibinfo{author}{{van Mulligen}, E.}, \bibinfo{author}{{Velterop}, J.},
  \bibinfo{author}{{Waagmeester}, A.}, \bibinfo{author}{{Wittenburg}, P.},
  \bibinfo{author}{{Wolstencroft}, K.}, \bibinfo{author}{{Zhao}, J.},
  \bibinfo{author}{{Mons}, B.}, \bibinfo{year}{2016}.
\newblock \bibinfo{title}{{The FAIR Guiding Principles for scientific data
  management and stewardship}}.
\newblock \bibinfo{journal}{Scientific Data} \bibinfo{volume}{3},
  \bibinfo{pages}{160018}.
\newblock \URLprefix \url{https://www.nature.com/articles/sdata201618},
  \DOIprefix\doi{10.1038/sdata.2016.18}.

\end{thebibliography}

\end{document}